\documentclass[9pt]{article}
\usepackage{spconf}
\usepackage{amsmath,graphicx, url}
\usepackage{subcaption}
\usepackage{booktabs}
\usepackage{adjustbox}
\usepackage{xcolor}
\usepackage{cite}
\usepackage{url}

\usepackage{breakurl}
\usepackage[breaklinks]{hyperref}

\def\NEWMODEL{{WARP-Q}}

\title{\NEWMODEL{}: Quality Prediction For Generative Neural Speech Codecs}
\name{Wissam A. Jassim$^1$, Jan Skoglund$^2$, Michael Chinen$^2$, Andrew Hines$^1$}
\address{
$^1$School of Computer Science, University College Dublin, Dublin, Ireland \\
$^2$Chrome Media, Google,  San Francisco, CA, USA\\
\\
\texttt{\small{wissam.a.jassim@gmail.com, jks@google.com, mchinen@google.com, andrew.hines@ucd.ie}}
}

\usepackage{tikz}
\usepackage{textcomp}
\usepackage{hyperref}
\usepackage{lipsum}

\newcommand\copyrighttext{%
  \footnotesize \textcopyright 2021 IEEE. Personal use of this material is permitted. Permission from IEEE must be obtained for all other uses, in any current or future media, including reprinting/republishing this material for advertising or promotional purposes, creating new collective works, for resale or redistribution to servers or lists, or reuse of any copyrighted component of this work in other works.}
\newcommand\copyrightnotice{%
\begin{tikzpicture}[remember picture,overlay]
\node[anchor=south,yshift=10pt] at (current page.south) {\fbox{\parbox{\dimexpr\textwidth-\fboxsep-\fboxrule\relax}{\copyrighttext}}};
\end{tikzpicture}%
}
    
\begin{document}
\ninept
\maketitle
\copyrightnotice

\begin{abstract}

Good speech quality has been achieved using waveform matching and parametric reconstruction coders. Recently developed very low bit rate generative codecs can reconstruct high quality wideband speech with bit streams less than 3 kb/s. These codecs use a DNN with parametric input to synthesise high quality speech outputs. Existing objective speech quality models (e.g., POLQA, ViSQOL) do not accurately predict the quality of coded speech from these generative models underestimating quality due to signal differences not highlighted in subjective listening tests. We present WARP-Q, a full-reference objective speech quality metric that uses dynamic time warping cost for MFCC speech representations. It is robust to small perceptual signal changes. Evaluation using waveform matching, parametric and generative neural vocoder based codecs as well as channel and environmental noise shows that WARP-Q has better correlation and codec quality ranking for novel codecs compared to traditional metrics in addition to versatility for general quality assessment scenarios.

\end{abstract}

\section{Introduction}
\label{sec:intro}
Estimation of speech quality is important for monitoring, evaluating, and developing speech transmission and communication services. Usually, assessing speech quality can be done subjectively with listening tests providing accurate results with small confidence intervals~\cite{6021874}. However, subjective methods are expensive and time consuming. Objective speech quality estimation models can provide a practical and efficient alternative for evaluating and predicting speech quality.

Different kinds of objective models exist depending on the speech applications and services. Models such as POLQA~\cite{polqa3},  PESQ~\cite{pesq_itu}, and ViSQOL~\cite{visqolv3, hines2015visqol} have been shown to work well for a wide variety of coding, channel and environmental degradations to the speech signal. They are full-reference (FR) metrics that compare a clean reference to a test signal that has been degraded. They do this by aligning and comparing the signals and mapping an estimate of the differences between the signal to a perceptual mean opinion score (MOS) scale.

Recently, data driven algorithms based on deep neural networks (DNNs) have created a new generation of generative speech synthesis models~\cite{45774, 8461368, DBLP:journals/corr/ArikDGMPPRZ17, prenger2019waveglow},
often with text-to-speech as the application. Of these, the auto-regressive teacher-forced architecture in WaveNet~\cite{45774}, WaveRNN~\cite{DBLP:journals/corr/abs-1802-08435}, and SampleRNN~\cite{mehri2016samplernn} has been used as the basis in new generative codecs \cite{Kleijn2017WavenetBL, 8682804, klejsa2019}. These codecs are wideband and are designed to operate at low bit rates, 
and have shown very promising results. The reconstructed audio  waveforms are generated by a neural network conditioned by traditional low bit rate parametric vocoder parameters, i.e., the speech signal is represented and transmitted as a sequence of parameters extracted at the encoder~\cite{Kleijn2017WavenetBL}.

The DNN-based codecs are generative, meaning that while the original and coded speech signals may both sound good, they have structural differences in them. This is because the decoded signal is generated from the parameters of the coded signal and from the model, so it may not fully align spectro-temporally with the original signal. These alignment differences cause problems for full reference quality models. Although they deal well with macro mis-alignments (delays etc.), micro-alignments across time or frequency components of speech cause quality prediction issues.

Although most common quality metrics such as POLQA and VISQOL metrics provide accurate quality scores for the speech signals processed by telephony and voice over IP (VoIP) transmission systems, they fail to provide acceptable results when speech signals are distorted by the effects of low bit rate DNN-based codecs. This work sought to develop a new FR model for speech quality prediction that worked for generative speech codecs. The aim was to have a general model that would work for speech from both generative and traditional codecs.

\section{Background}


Speech codec algorithms are designed to compress speech signals at a low bit rate and yet retaining high speech quality. The processes include analysing and converting spoken sounds into digital codes and vice versa. Over decades, different types of codecs have been introduced~\cite{10.5555/546671}. Different aspects such as bit rate, language of spoken words, channel errors, coding delays, and memory and computational cost determine the need and performance of any speech codec algorithm. 

Speech coders are traditionally classified into two types of algorithms: parametric, such as MELP~\cite{540325} and waveform (matching) codecs, e.g., Speex~\cite{speex2007} and Opus~\cite{rfc6716}. Traditional parametric codecs are generative in that they extract speech features controlling a generative synthesis, focusing on sounding similar to the input and disregarding the actual output waveform. While this can cause fidelity issues it does not significantly impact time alignment. The WaveNet~\cite{45774,Kleijn2017WavenetBL} and LPCNet~\cite{8682804, Valin2019} codecs are examples of the new class of generative neural codecs, where the synthesis is driven by a neural network. These neural-based generative codecs can produce high fidelity output but pose alignment challenges to objective quality metrics.

We previously compared the performance of these two types of codecs in terms of several quality aspects, such as accuracy of pitch periods estimation, the word error rates for automatic speech recognition (ASR), and the influence of speaker gender and coding delays~\cite{Wissam2020}. It was observed that these factors should be taken into account in order to design a new and robust FR metric that is workable for different codecs. We analysed why existing speech quality models underrated the quality of generative codec outputs and considered the micro-alignment differences as a potential cause.

Generative codecs rely on a combination of parametrically coded information and a neural model (e.g., WaveNet) to synthesise the output. Although the resulting codec speech is rated as high quality, there are small differences between the original and codec speech. Some of these are temporal micro-alignments and others manifest as slight pitch shifts. While these are potentially perceptible differences, a human listener may not be able to distinguish a quality difference. On the other hand, traditional codecs keep  the original and coded signals temporally aligned and may be penalised in subjective ratings due to the spectro-temporal differences that manifest as noise or corrupt speech in the coded output. 

Standard speech quality models rely on evaluating the \textit{similarity} between reference and test signals as a salient feature for assessing quality. They pre-align the signals in order to account for quality issues resulting from delay and signal corruption. For example, the ViSQOL metric~\cite{visqolv3,hines2015visqol} uses the neurogram similarity index measure (NSIM) to estimate the similarity between a pre-aligned reference patch and a degraded spectrogram patch frame by frame. 

In this study, we propose \NEWMODEL{}, based on dynamic time warping (DTW), calculating an optimal match between two given sequences. DTW has been successfully adopted for a range of speech processing applications. In~\cite{Kraljevski2008PerceivedSQ}, the \textit{global} alignment distance based on the \textit{original} DTW is employed for test and received speech comparison. It showed results comparable to that of the PESQ metric for perceived speech quality measurement in VoIP and global system for mobile communications (GSM) networks.  

\NEWMODEL{} takes a different approach to traditional speech quality models handling time-alignment and signal similarity in a combined manner. We use a special type of DTW algorithm, known as \textit{subsequence} dynamic time warping (SDTW)~\cite{10.5555/2815664}, to measure the \textit{distance} between speech signals. Unlike the original DTW algorithm which aims to find an optimal global alignment between two given sequences, the SDTW finds a subsequence within the longer sequence that optimally fits the shorter sequence. It has been successfully employed in audio matching scenarios and content-based audio retrieval 
applications~\cite{4523006}.
We refer to Figs. 3.10 and 7.13 from M\"{u}ller~\cite{10.5555/2815664} for more details about the difference between the original DTW and SDTW.  


We show that this simple concept allows the perceptual quality impact of micro-alignment and signal corruption to be captured and quantified together. The proposed SDTW-based metric predicts speech quality correctly for generative codecs while also performing competitively with standard metrics for a wide range of standard coding algorithms and distortion effects.  

\vspace{-3mm}
\section{Proposed Algorithm}
\label{sec:format}
\begin{figure}[th!]
  \centering
  \centerline{\includegraphics[width=9cm]{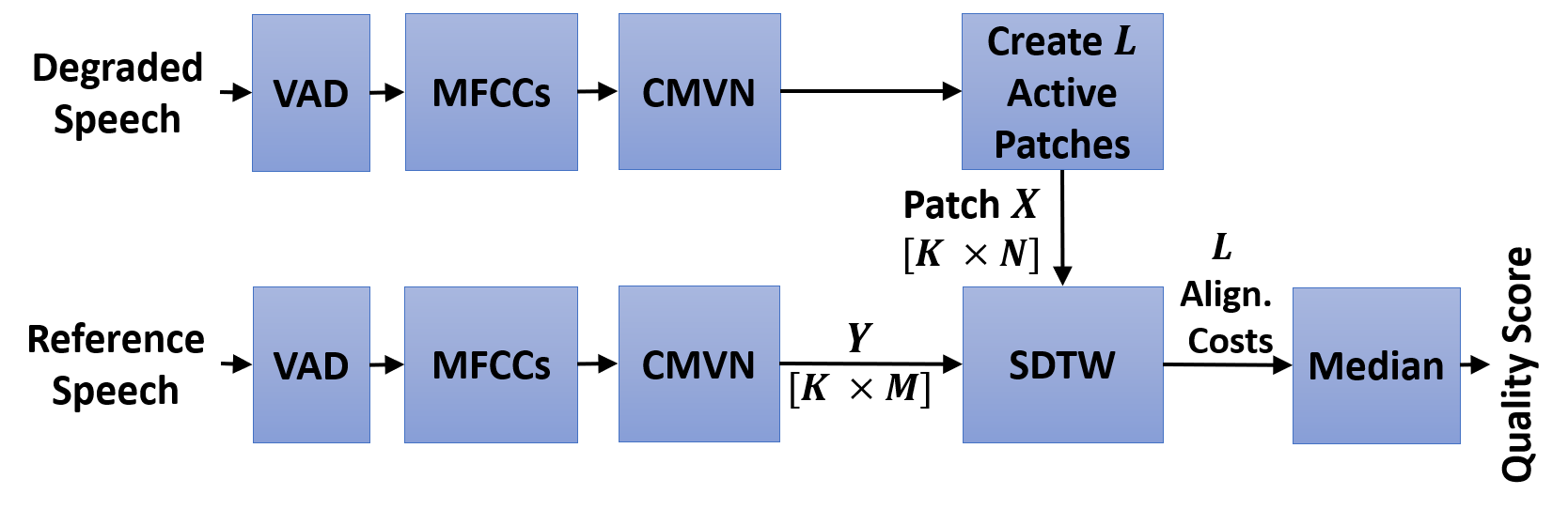}}
\caption{Block diagram of the proposed \NEWMODEL{} metric. 
}
\label{fig1}
\end{figure}

Fig.~\ref{fig1} illustrates the four processing stages of the proposed algorithm: pre-processing; feature extraction; similarity comparison; and subsequence score aggregation. Python source code for the \NEWMODEL{} model is available for download 
in~\cite{warpq_code}.
 

\subsection{Pre-processing}
The reference and degraded input signals are set to the same sampling frequency, $f_s=16$~kHz. Silent non-speech segments are removed from reference and degraded signals using a voice activity detection (VAD) algorithm. Our implementation used a WebRTC-based VAD with default parameters~\cite{webRTCVAD}.       

\subsection{Spectral Features}
Mel frequency cepstral coefficients (MFCCs) representations of the reference and degraded signals are generated using 12 critical bands up to 5~kHz for each frame~\cite{mcfee2015librosa}. A Hann window with a length of 32~ms and 80$\%$ overlap was used for framing. Spectral coefficients were extracted using the discrete cosine transform (DCT) type-2 with orthonormal bases and a cepstral filtering of 3 for liftering. 
The MFCCs signal representations are normalised so that they have the same segmental statistics (zero mean and unit variance). The spectral coefficients of each feature vector were normalised using the cepstral mean and variance normalisation (CMVN) algorithm~\cite{amirsina_torfi_2017_840395}. 

\subsection{Subsequence dynamic time warping (SDTW)}
Let $X=(x_1,x_2,...,x_N)$ and $Y=(y_1,y_2,...,y_M)$ be two feature sequences over a feature space. The length $M$ is assumed to be much larger than the length $N$. For the two given sequences, the SDTW algorithm considers all possible subsequences of $Y$ to find the optimal one that minimises the DTW distance to $X$. The optimal subsequence of $Y$ is determined by two optimal indices $a^{\ast}$ and $b^{\ast}$, where $a^{\ast}, b^{\ast}\in [1\colon M]$ with $a^{\ast} \leq b^{\ast}$, such that the subsequence $Y(a^{\ast} \colon b^{\ast})$ has the minimum DTW distance to $X$ over other subsequences. To reveal the optimal index $b^{\ast}$, the algorithm computes the $N\times M$ accumulated cost matrix denoted by $D$ using dynamic programming. The index that minimises cost values in \textit{last} row of $D$ represents the optimal index $b^{\ast}$. To reveal the optimal index $a^{\ast}$, the algorithm drives the optimal warping path $P^\ast$ (list of index pairs) between $X$ and $Y(a^{\ast} \colon b^{\ast})$ using backtracking, which starts with $q_1 = (N,b^{\ast})$ and stops as soon as the \textit{first} row of $D$ is reached by some element $q_r = (1,m), m \in [1 \colon M]$. Refer to Exercise 7.6 from M\"{u}ller, 2015~\cite{10.5555/2815664} for more details about the SDTW algorithm. We use the Librosa Python library~\cite{mcfee2015librosa} to implement the SDTW algorithm with a step size condition of $\sum\{(1, 1), (3, 2), (1, 3)\}$ and an Euclidean cost function. 


\begin{figure*}[ht]
\begin{minipage}[b]{0.33\linewidth}
  \centering
  \centerline{\includegraphics[width=6.0cm]{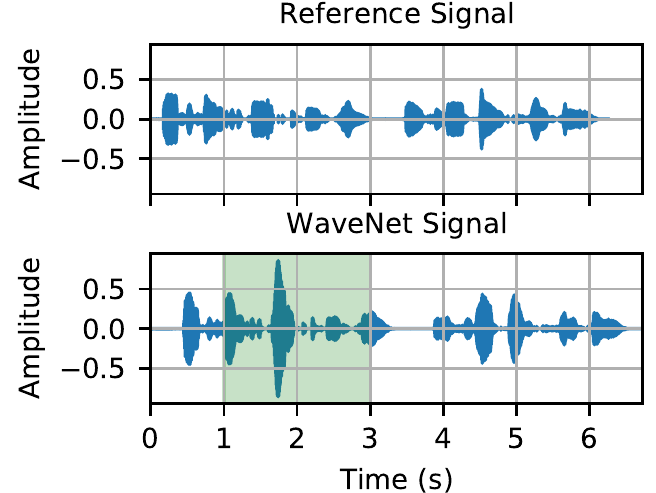}}
  \subcaption{\textsl{Signals in time space}\label{fig2_a}}
\end{minipage}
\begin{minipage}[b]{0.33\linewidth}
  \centering
  \centerline{\includegraphics[width=6.0cm]{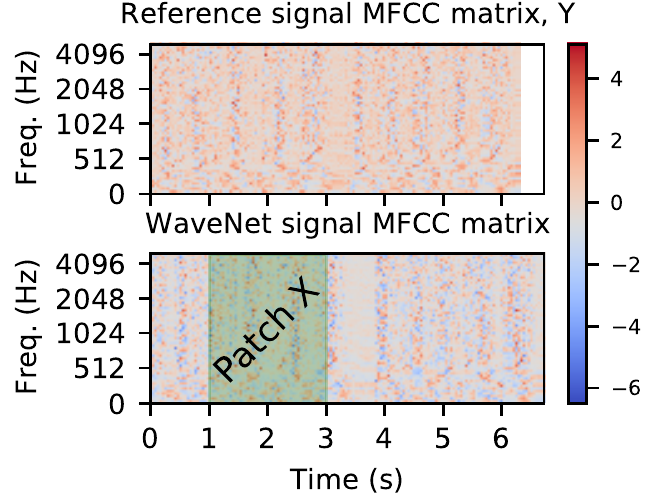}}
  \subcaption{\textsl{MFCC feature space}\label{fig2_b}}
\end{minipage}
\begin{minipage}[b]{0.33\linewidth}
  \centering
  \centerline{\includegraphics[width=6.0cm]{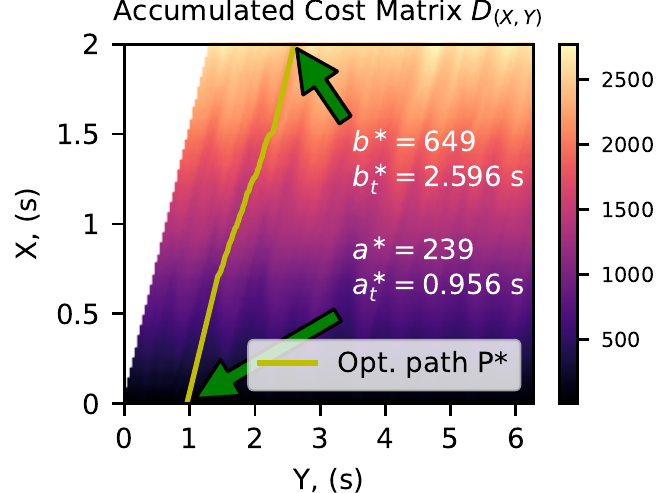}}
  \subcaption{\textsl{Sub-signal DTW space}\label{fig2_c}}
\end{minipage}

\caption{SDTW-based accumulated cost and optimal path between two signals. (a) plots of a reference signal and its corresponding coded version from a WaveNet coder at 6 kb/s (obtained from the VAD stage), (b) normalised MFCC matrices of the two signals, (c) plots of SDTW-based accumulated alignment cost matrix $D_{(X,Y)}$ and its optimal path $P^\ast$ between the MFCC matrix $Y$ of the reference signal and a patch $X$ extracted from the MFCC matrix of the degraded signal. The optimal indices ($a^{\ast} \& b^{\ast}$) are also shown. $X$ corresponds to a short segment (2 s long) from the WaveNet signal (highlighted in green color). }
\label{fig2}
\end{figure*}


\subsection{Quality score computation}
The reference and degraded MFCC representations can be treated as 2-dimensional matrices for processing. The reference MFCC matrix, $Y$, has a size of $K \times M$, where $K=12$ which represents the 12 frequency bands, and $M$ is the total number of signal frames. The MFCC matrix of degraded signal is divided into a number, $L$, of patches with a $50\%$ overlap. Each patch, $X_i$ ($i=1,2,...,L$), is of size $K \times N$, where $N=100$ corresponds to 100 frames (400 ms) patch length from the degraded speech. $L$ is equal to the total number of degraded signal frames divided by $N$. 
For each degraded patch $X_i$, the SDTW algorithm described above is used to compute the accumulated cost matrix $D_{(X_i,Y)}$, optimal warping path $P^\ast$, and optimal indices $a^{\ast}$ and $b^{\ast}$ between $X_i$ and the reference MFCC matrix $Y$. The accumulated cost of index $b^{\ast}$ is adopted as the salient feature for quality score estimation. The quality score per each degraded patch is computed as follows: 
\begin{equation}
C_i = \frac{1}{N} D_{(X_i,Y)}[N,b^{\ast}], \, \, i=1,2,...,L.
\label{eq_ci}
\end{equation}
Note that in Eq.~\ref{eq_ci}, the accumulated cost is divided by $N$ (length of degraded patch) to suppress the dynamic range of the predicted scores. Eq.~\ref{eq_ci} provides a vector of $L$ elements corresponding to the total number of patches in the degraded signal. Finally, the aggregate quality score (QS) for the degraded signal is computed as the median of the cost per patch:

\begin{equation}
QS = \text{Median}([C_1,C_2,...,C_L]).
\label{eq_qs}
\end{equation}

\begin{figure*}[t]
\begin{minipage}[b]{0.24\linewidth}
  \centering
  \centerline{\includegraphics[width=4.2cm]{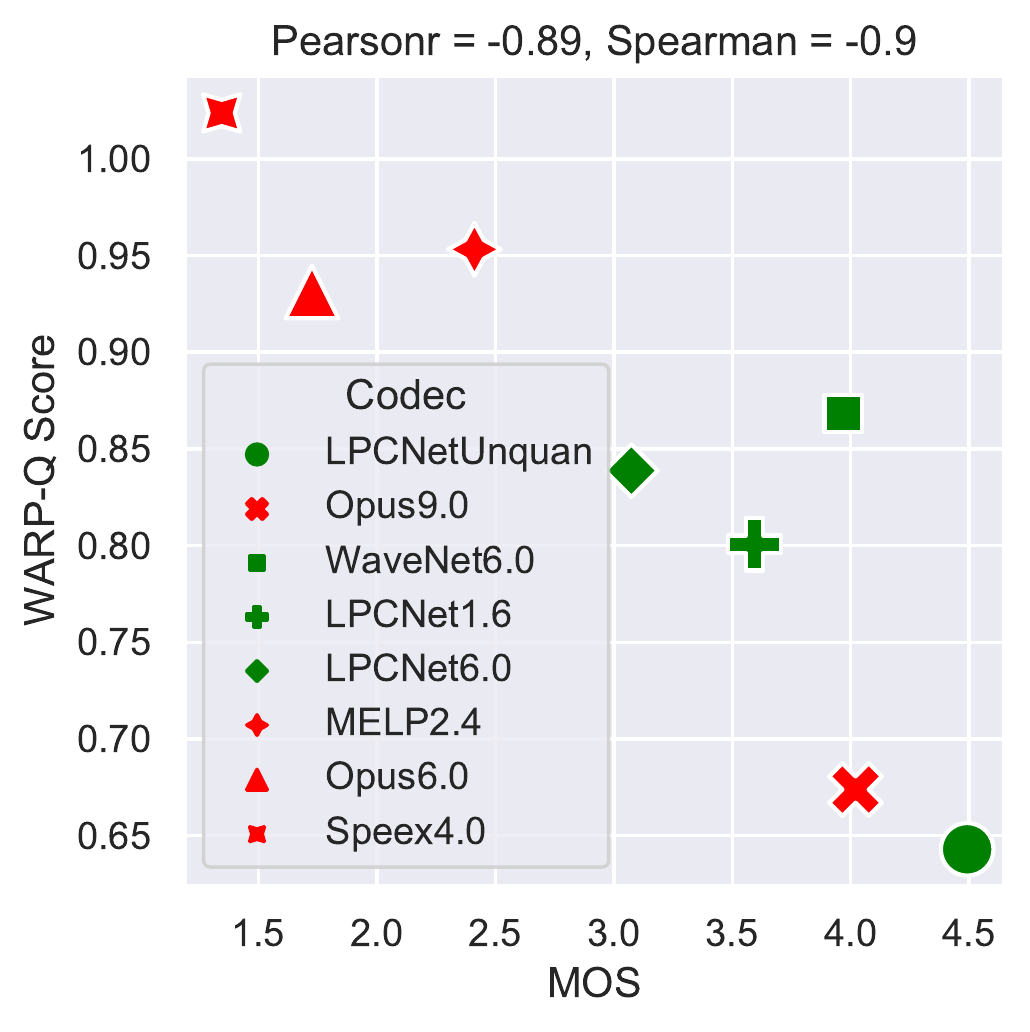}}
  \subcaption{\textsl{}\label{fig4_a}}
\end{minipage}
\begin{minipage}[b]{0.24\linewidth}
  \centering
  \centerline{\includegraphics[width=4.2cm]{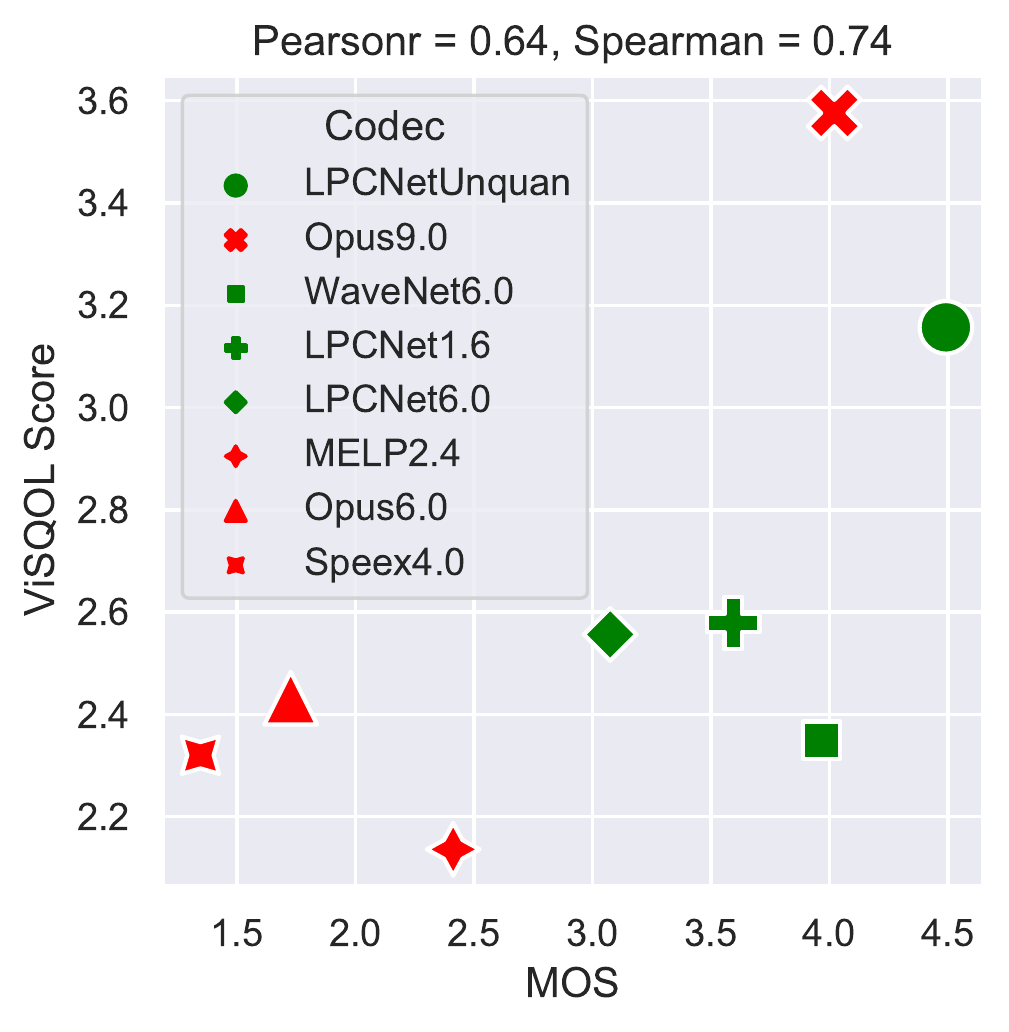}}
  \subcaption{\textsl{}\label{fig4_b}}
\end{minipage}
\begin{minipage}[b]{0.24\linewidth}
  \centering
  \centerline{\includegraphics[width=4.2cm]{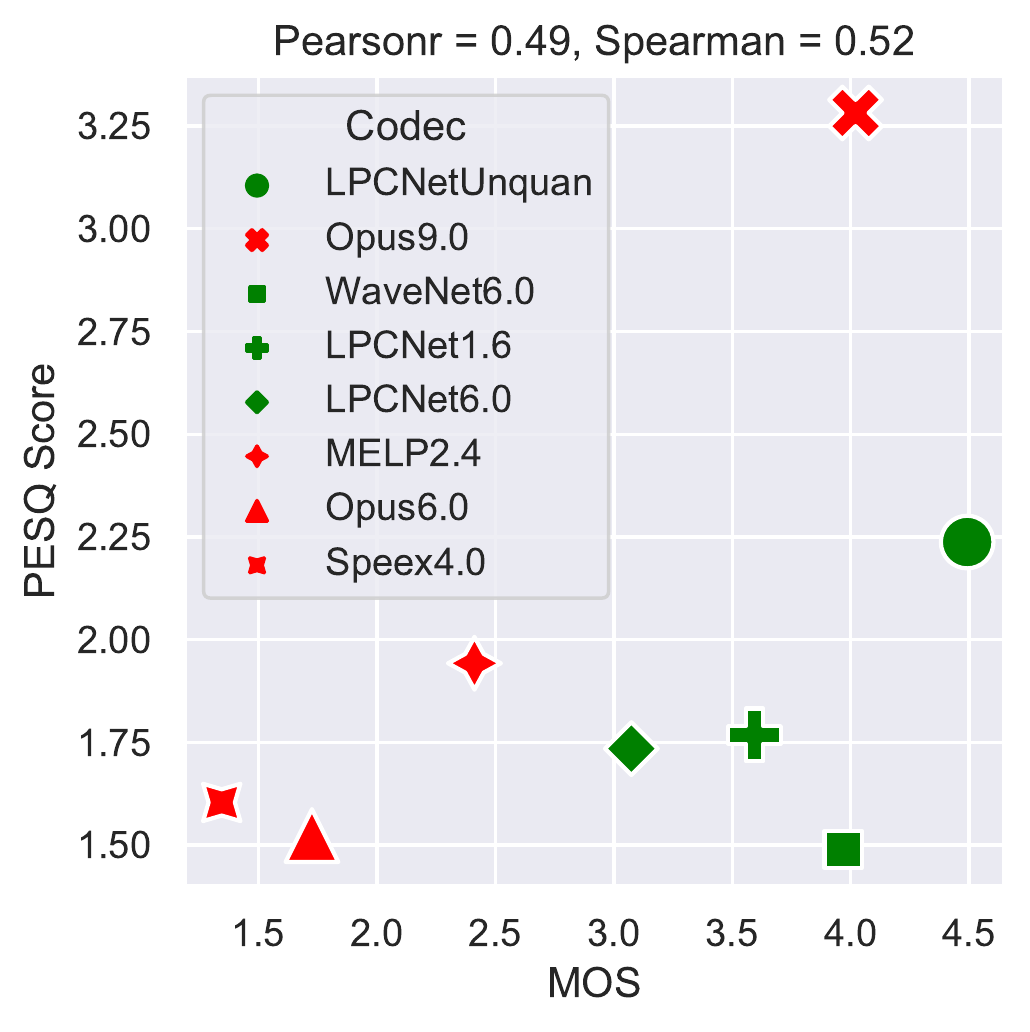}}
\subcaption{\textsl{}\label{fig4_d}}
\end{minipage}
\begin{minipage}[b]{0.24\linewidth}
  \centering
  \centerline{\includegraphics[width=4.2cm]{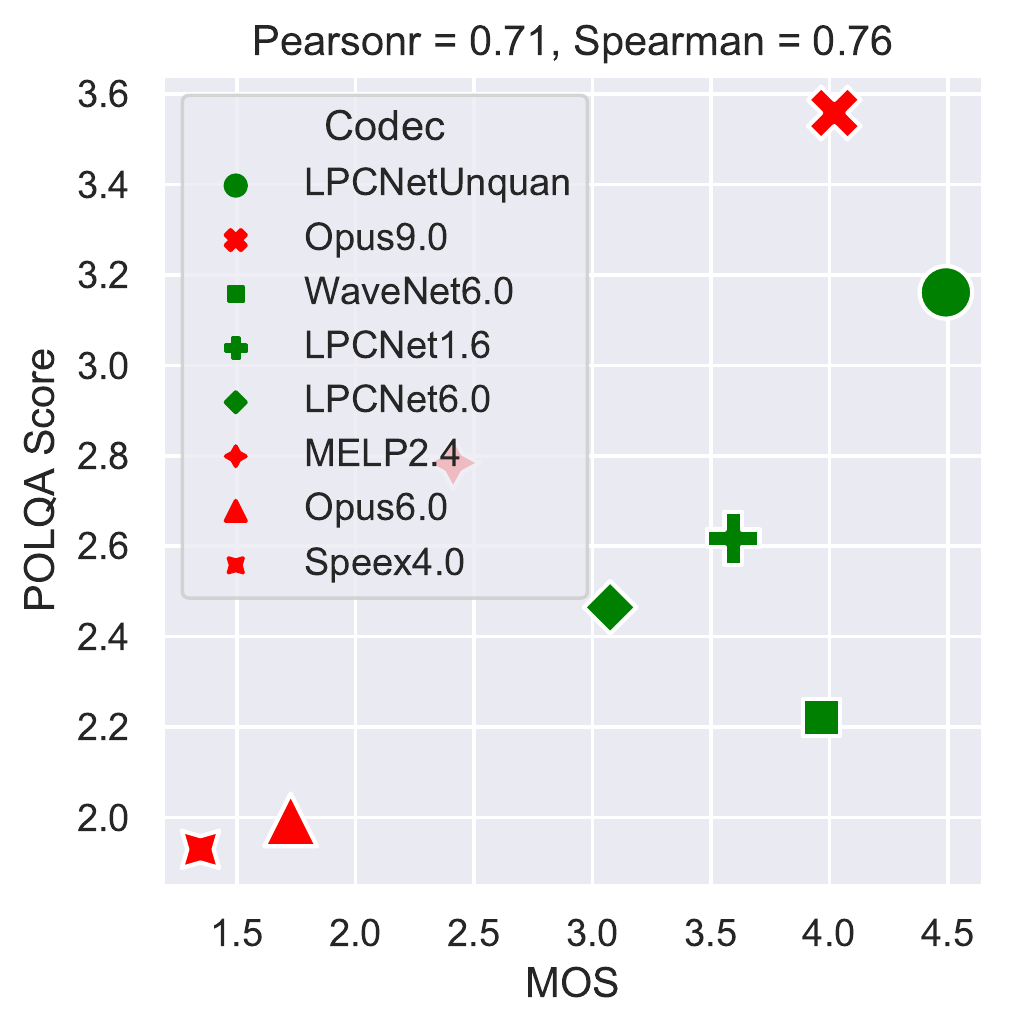}}
\subcaption{\textsl{}\label{fig4_f}}
\end{minipage}

\caption{Per condition QS predicted for the Genspeech database using: (a) the proposed metric, (b) ViSQOL metric, (c) PESQ metric, and (d) POLQA metric. Points are highlighted in two different colors: in green for generative neural coders (LPCNetUnquan, WaveNet6.0, LPCNet1.6 and LPCNet6.0) and in red for traditional coders (Opus9.0, MELP2.4, Opus6.0 and Speex4.0).}
\label{fig:genspeechresults}
\end{figure*}

\begin{figure*}[t]
\begin{minipage}[b]{0.24\linewidth}
  \centering
  \centerline{\includegraphics[width=4.2cm]{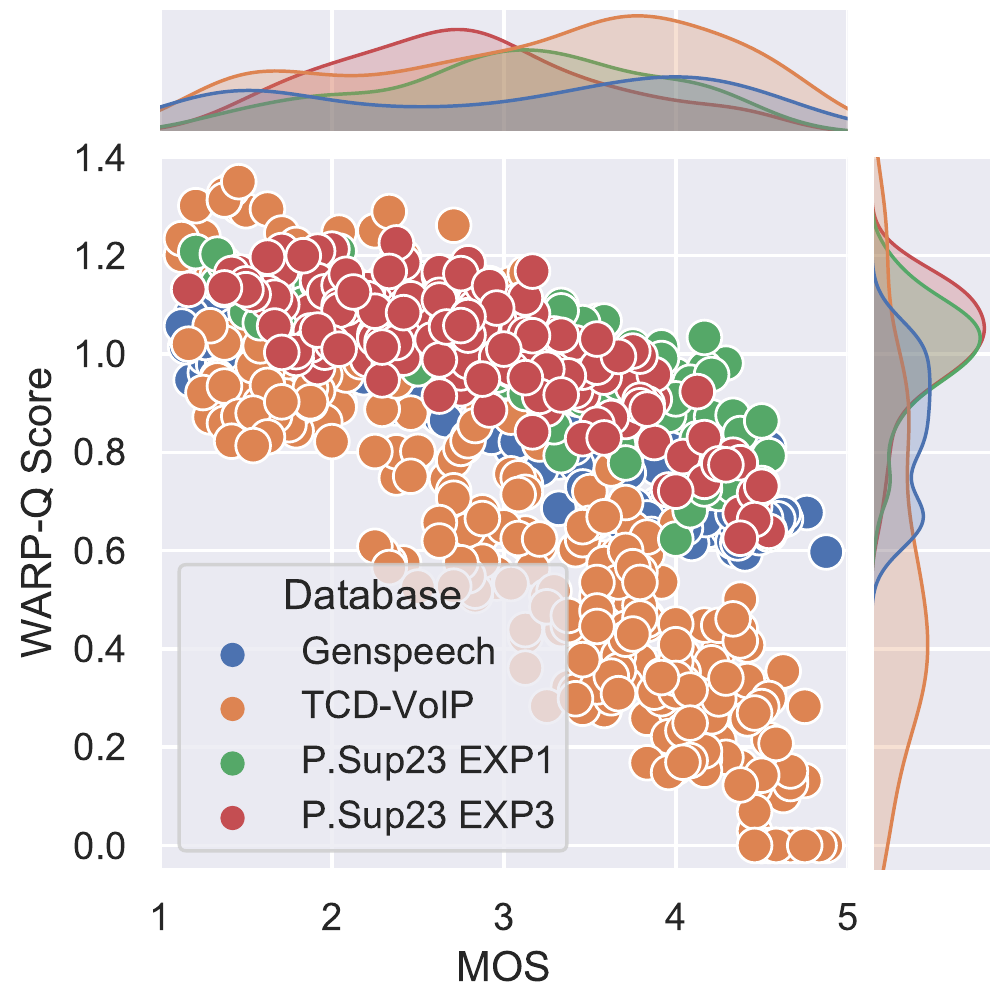}}
  \subcaption{\textsl{}\label{fig3_a}}
\end{minipage}
\begin{minipage}[b]{0.24\linewidth}
  \centering
  \centerline{\includegraphics[width=4.2cm]{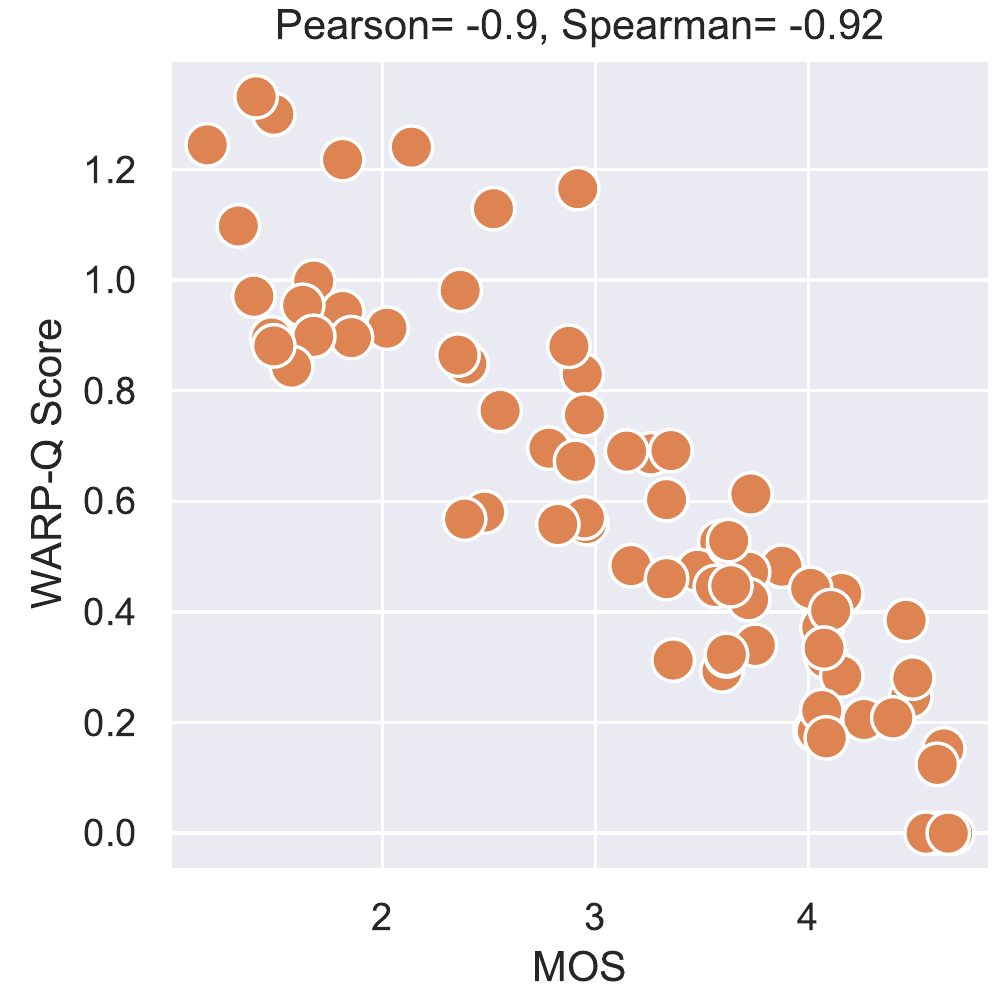}}
  \subcaption{\textsl{}\label{fig3_b}}
\end{minipage}
\begin{minipage}[b]{0.24\linewidth}
  \centering
  \centerline{\includegraphics[width=4.2cm]{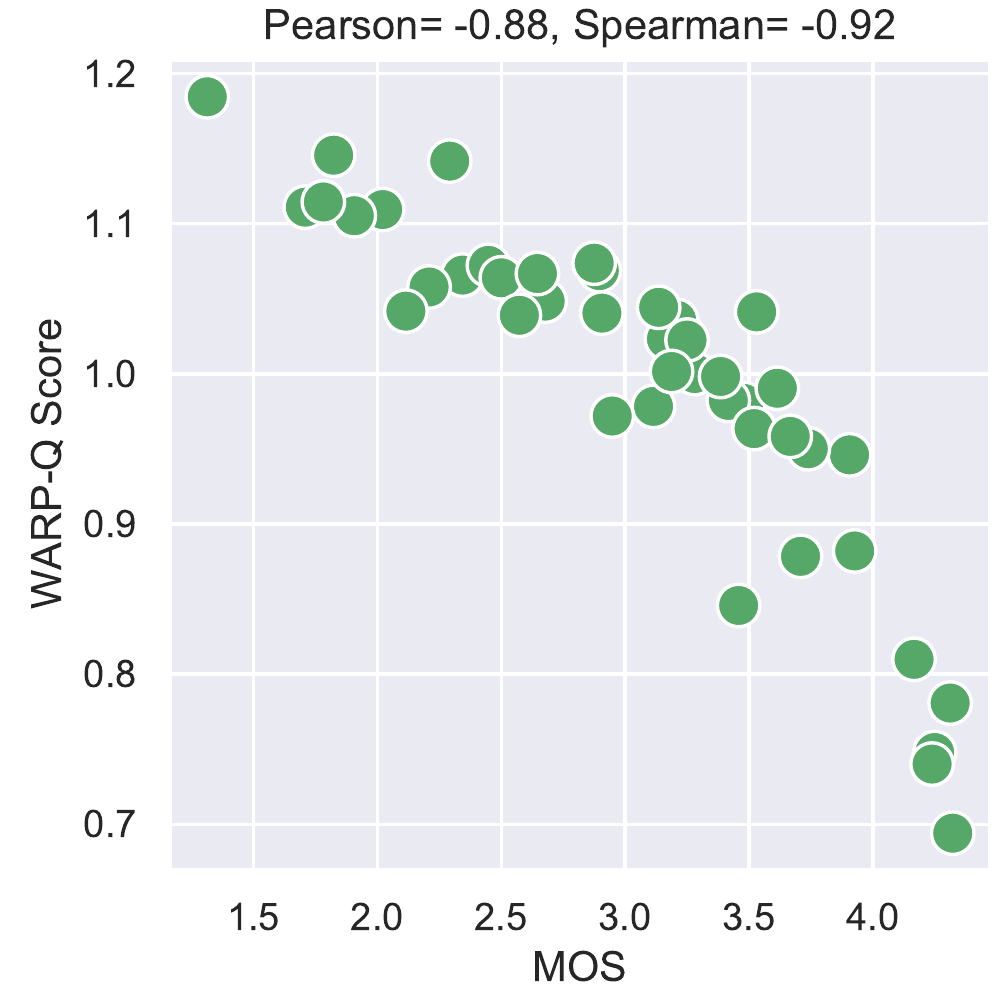}}
\subcaption{\textsl{}\label{fig3_c}}
\end{minipage}
\begin{minipage}[b]{0.24\linewidth}
  \centering
  \centerline{\includegraphics[width=4.2cm]{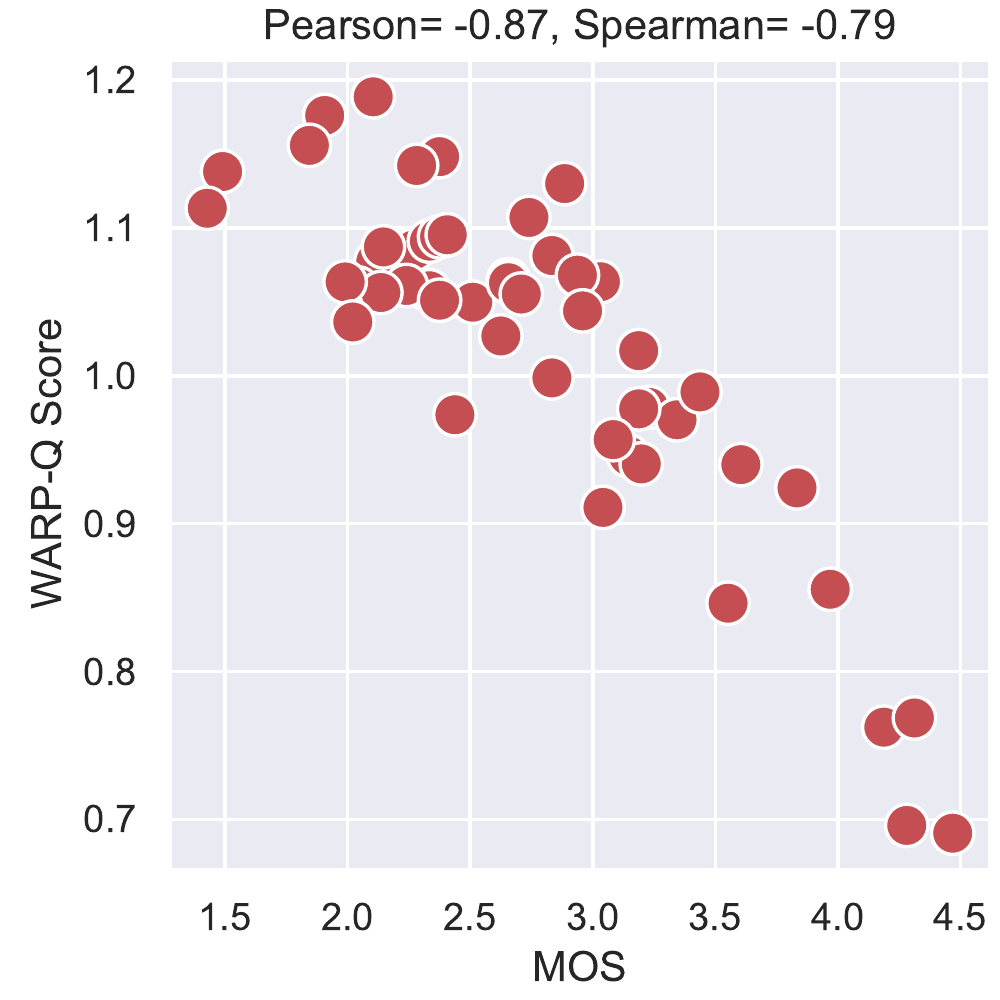}}
\subcaption{\textsl{}\label{fig3_d}}
\end{minipage}

\caption{QS predicted by \NEWMODEL{} using Eq.~\ref{eq_qs} for: (a) per sample scores for all databases, (b) per condition scores for the TCD-VoIP database, (c) per condition scores for the P.Sup23 EXP1 database, and (d) per condition scores for the P.Sup23 EXP3 database. }
\label{fig:allscatter}
\end{figure*}

An illustration of the process is presented in Fig.~\ref{fig2}. A reference signal taken from the Genspeech database \cite{Wissam2020} with its corresponding coded version from a WaveNet coder at 6 kb/s are fed to the VAD algorithm to remove non-speech segments from them. The plots of the two processed signals are shown in Fig.~\ref{fig2_a}. Their normalised MFCC representations are shown in Fig.~\ref{fig2_b}. A patch $X$ is extracted from the MFCC array of the degraded signal. In this example, for better visualisation, we used a wider patch of 2 s length ($N=500$, which corresponds to 500 frames long). 
The extracted patch is highlighted in green color in Figs.~\ref{fig2_a} and ~\ref{fig2_b}. Fig.~\ref{fig2_c} displays $D_{(X,Y)}$ with its corresponding $P^\ast$ computed by the SDTW algorithm. The optimal index located in the top row of $D_{(X,Y)}$ is $b^{\ast}=649$, which corresponds to $649*4 \text{ms}/\text{frame}=2.596$~s in time, i.e.,  $b^{\ast}_{t}=2.596$ s. Furthermore, the optimal index located in the bottom row of $D_{(X,Y)}$ is $a^{\ast}=239$, which corresponds to $0.956$~s time index, i.e., $a^{\ast}_{t}=0.956$~s. This indicates that the subsequence of $Y$ that has the minimum alignment cost distance to patch $X$ is $Y(a^{\ast}=239 \colon b^{\ast}=649)$, i.e., $Y(a^{\ast}_{t}=0.956 \colon b^{\ast}_{t}=2.596)$~s in time.

\section{Experimental Evaluation}
\label{Evaluating_algorithm}
Data from subjective experiments on a variety of parametric and generative codecs (the Genspeech database, with data from \cite{Valin2019,Skoglund2020ImprovingOL}) are used to evaluate \NEWMODEL{} and benchmark the performance against existing models. Table~\ref{tab:genspeech} summarises the codecs (further details available in~\cite{Wissam2020}). Furthermore, the capability of \NEWMODEL{} to predict speech quality for quality issues beyond low bit rate coding is evaluated using other datasets: the TCD-VoIP~\cite{TCDVOIP}, a database which contains speech signals under a range of common VoIP degradations with channel and environmental issues, and the ITU-T P. Supplement 23 (P.Sup23) [EXP1 and EXP3]~\cite{itu_sup23}, a database which contains speech samples under a range of traditional coding and some environmental degradations. Note that the original MUSHRA scores from the Genspeech database were linearly rescaled to be in the same range of MOS of other databases.   

\begin{table}[tbp!]
  \centering
 \caption{The Genspeech dataset. Further details in \cite{Wissam2020}.}
 \label{tab:genspeech}
    \begin{adjustbox}{width=\columnwidth}
\begin{tabular}{llll}
Codec & Bit rate & Description \\
\hline
LPCNetUnquant & --- & LPCNet operating on Opus unquantized features \\
Opus9.0 & 9 kb/s & Wideband vocoder (SILK mode) \\
WaveNet6.0 & 6 kb/s & WaveNet operating on Opus quantized features \\
LPCNet1.6 & 1.6 kb/s & WaveRNN + linear prediction \\
LPCNet6.0 & 6 kb/s & LPCNet operating on Opus quantized features \\
MELP2.4 & 2.4 kb/s & Narrowband vocoder \\
Opus6.0 & 6 kb/s & Narrowband vocoder (SILK mode) \\
Speex4.0 & 4 kb/s & Wideband vocoder (wideband quality 0) \\
\end{tabular}
\end{adjustbox}
\end{table}

Models are compared using Pearson’s correlation coefficient and Spearman rank-order correlation coefficient. Fig.~\ref{fig:genspeechresults} compares the per condition (i.e., grouped by codec) \NEWMODEL{} scores to that of existing metrics for the Genspeech dataset. The proposed metric provided scores that are ranked and consistent more than other metrics for all codecs. 

Fig.~\ref{fig3_a} presents a scatter plot of \NEWMODEL{} scores against subjective quality ratings for samples from the four datasets at a standardised sampling frequency ($f_s=16$ kHz). A consistent inverse correlation between \NEWMODEL{} scores and MOS is apparent and the range of predicted scores is good between datasets. This highlights the robustness of the proposed algorithm to different degradation scenarios as the predicted quality scores remain bounded in a similar range. 
Fig.~\ref{fig3_b}-\ref{fig3_d} present the promising per condition results predicted by~\NEWMODEL{} for the TCD-VoIP and P.Sup23 [EXP1 and EXP3] datasets.

Table~\ref{tab_all_metics} presents correlation statistics per condition, for each metric by dataset. The proposed metric shows promise as a general use speech quality model for coding, channel, noise and other quality degradations in a competitive way to the PESQ, POLQA, and ViSQOL metrics.

\begin{table}[tbp]
  \centering
  \caption{Benchmark Statistics}
  \begin{adjustbox}{width=\columnwidth}
    {
    \begin{tabular}{lrrrr}
    Database: & \multicolumn{1}{l}{Genspeech} & \multicolumn{1}{l}{TCD-VoIP} & \multicolumn{1}{l}{P.Sup23 EXP1} & \multicolumn{1}{l}{P.Sup23 EXP3} \\
\cmidrule{2-5}          & \multicolumn{4}{c}{Pearson} \\
    \NEWMODEL{} & \textbf{-0.89} & -0.9  & -0.88 & -0.87 \\
    ViSQOL & 0.64  & 0.74  & 0.87  & 0.78 \\
    PESQ & 0.49  & \textbf{0.91} & 0.91  & 0.87 \\
    POLQA & 0.71  & 0.89  & \textbf{0.96} & \textbf{0.96} \\
          & \multicolumn{4}{c}{Spearman} \\
    \NEWMODEL{} & \textbf{-0.9} & \textbf{-0.92} & -0.92 & -0.79 \\
    ViSQOL & 0.74  & 0.76  & 0.89  & 0.67 \\
    PESQ & 0.52  & 0.91  & 0.96  & 0.87 \\
    POLQA & 0.76  & 0.89  & \textbf{0.97} & \textbf{0.94} \\
    \end{tabular}}
    \end{adjustbox}
  \label{tab_all_metics}%
\end{table}%

\vspace{-3 mm}

\section{Discussion and Conclusions}
\label{conclusions}
Generative coding is changing the fundamental relationship between the source and codec signal: if the reference and codecs signal are perceptually different in timing or pitch but indistinguishable in quality then subjectively they are both high quality. Traditional objective quality metrics have matured over the last two decades but do not easily adapt to the challenge posed by generative codecs. A new approach is needed. Adopting the SDTW algorithm and applying it to MFCC features allows \NEWMODEL{} to be resilient to micro-alignment issues while penalising perceptible signal intensity changes caused by coding artefacts. The results show that although \NEWMODEL{} is a simple model building on well established speech signal processing features and algorithms it solves the unmet need of a speech quality model that can be applied to generative neural codecs. Work is ongoing to further optimise the model (e.g. DTW parameters, choice of MFCC representation, median aggregation) and add a cognitive mapping subsystem to map \NEWMODEL{} scores to a human subjective (MOS) rating scale. Python code of WARP-Q is available on GitHub for ease of access and contribution.  

\section{Acknowledgment}
The authors would like to thank Google for kindly providing the subjective labelled data of low bit rate codecs, ViSQOL, PESQ, and POLQA quality scores. This work has emanated from research supported in part by the Google Chrome University Program and research grants from Science Foundation Ireland (SFI) co-funded under the European Regional Development Fund under Grant Number 13/RC/2289\_P2 and 13/RC/2077.

\vfill
\bibliographystyle{IEEEtran}
\bibliography{main}
\vfill
\end{document}